**Impact of Nitrogen incorporation on pseudomorphic site-controlled quantum dots grown by Metalorganic Vapour Phase Epitaxy**

V. Dimastrodonato, L.O. Mereni, G. Juska, and E. Pelucchi

*Epitaxy and Physics of Nanostructures, Tyndall National Institute, University College Cork, Cork, Ireland*

**Abstract**

We report on some surprising optical properties of diluted nitride $InGaAs_{1-\varepsilon}N_\varepsilon/GaAs$ ($\varepsilon<<1$) pyramidal site-controlled Quantum Dots, grown by Metalorganic Vapour Phase Epitaxy on patterned GaAs (111)B substrates. Micro-photoluminescence characterizations showed anti-binding exciton/bi-exciton behaviour, a spread of exciton lifetimes in an otherwise very uniform sample, with unexpected long neutral exciton lifetimes (up to 7ns) and a nearly-zero Fine Structure Splitting on a majority of dots.

In the last decades researches in fundamental physics, electrodynamics, and the more recently arisen technologies of quantum information and quantum computing have regarded semiconductor Quantum Dots (QDs) as ideal candidate for both theoretical studies and practical applications. The interest QDs have attracted is amplified by the possibility of controlling and manipulating their electro-optical properties. Emission energies can be tuned by modifying the dot dimensions and/or the dot-barriers band discontinuity.[1] A fine tuning of the exciton/bi-exciton binding energy can be achieved through a size scalability mechanism or via the application of electro-magnetic fields.[2] Strain induced, electric and magnetic fields can be used to reduce the built-in asymmetries which produce a Fine Structure Splitting (FSS) in the cascade bi-exciton/exciton pair, critical for a reliable source of entangled photons.[3]

More recent investigations on self-assembled In(Ga)As/GaAs QDs have suggested tuning mechanisms of the dot properties through strain engineering. [4] Employing strain-controlled schemes in these systems would overcome the main limiting factor which still inhibits a further redshift of the emission wavelength. The compressive strain effect, originating from the high difference between GaAs substrate and In(Ga)As dot layer lattice constant, limits in fact an increase of the thickness and/or Indium concentration. Moreover, the asymmetric strain distribution is one of the causes of the lack of degeneracy of the cascade bi-exciton/exciton photon pair, hampering the generation of polarized entangled photons.

It has been demonstrated that a small quantity of Nitrogen into (In)GaAs layers can partially compensate the compressive strain,[5] reducing relaxation processes and the consequent degradation of the emission properties.

Incorporating Nitrogen remains though a real issue, both for Molecular Beam Epitaxy [6] and Metalorganic Vapour Phase Epitaxy (MOVPE) techniques: the actual low incorporation efficiency requires particular growth conditions, based mainly on low temperatures and low group V source partial pressures,[7] which lead to a generally poorer crystallographic quality of the grown structures.

We present in this letter unforeseen optical properties of pseudomorphic pyramidal site controlled $InGaAs_{1-\varepsilon}N_\varepsilon/GaAs$ QDs grown by MOVPE on (111)B GaAs substrates. These systems have shown both high uniformity and spectral purity on par with self assembled samples. Recently a simple wavelength tuning mechanism was achieved by changing the dot Indium concentration.[8,9] The strain in the InGaAs layers nevertheless limited the QD emission wavelength to ~880nm (for a nominal Indium concentration of 45%) and delivered poorer spectral quality in the samples with higher In content.

To limit these deteriorating effects, we started growing and investigating diluted nitride QD systems. We report here on the properties of a specific sample structure (fully reproducible in its properties) in the limit of ultra nitrogen dilution. We anticipate that, although no substantial redshift was observed in the emission spectra from the specimen here analyzed (for other InGaAsN/GaAs systems, grown under different conditions, we have a more significant increase of the emission wavelength), the small concentration of Nitrogen incorporated in the dot changed other optical features of the system: an unexpected anti-binding bi-exciton/exciton behaviour was observed and a significant percentage of neutral excitons show emission decay times up to 7ns, while most of them exhibit lifetimes in the range 1ns-4ns. Moreover a FSS remarkably lower than precedent Nitrogen free InGaAs/GaAs dots was measured.

Diluted nitride InGaAs$_{1-\varepsilon}$N$_\varepsilon$ QDs were grown by MOVPE with standard metalorganic sources into 7.5µm pitch pyramidal recesses, chemically etched onto GaAs (111)B substrates and acting as nucleation seeds during the growth of the dot structure.[10] The sample design is similar to what published in Ref. 9, with the only difference being the 0.5 nm thick In$_{0.25}$Ga$_{0.75}$As$_{1-\varepsilon}$N$_\varepsilon$ dot layer embedded between GaAs barriers. The QD layer was grown at 730°C, with an AsH$_3$/III ratio of 750 and a flux ratio U-DMHydrizine /AsH$_3$ of 4/3.

Although state of art diluted nitride QD systems are generally grown at relatively low temperatures[7] to ease the Nitrogen incorporation,[11] our findings suggest that, as it will be clear in the subsequent text, a low percentage of Nitrogen can be incorporated into our dot systems at temperatures as high as 730°C. Further studies with the goal to better understand the dependence of the Nitrogen incorporation on the temperature (and V/III ratios) are necessary and ongoing on other specimens.

In Fig.1 the layer sequence of a typical pyramidal QD structure is shown through an Atomic Force Microscopy (AFM) image of a representative 1.5 nm thick In$_{0.35}$Ga$_{0.65}$As/GaAs dot, cleaved along one of the (110) directions and scanned in cross section. The dot, buried along the central axis of the inverted pyramid, is labelled in the same figure. The contrast between AlGaAs and pure GaAs is given by the presence of a thin oxide layer on AlGaAs alloy, while the stress released after cleaving the sample between InGaAs and the confining GaAs makes the dot layer identification clear.

After growth a processing known as surface etching was performed on the sample in order to chemically remove irregular surface faceting and nanoclusters whose emission can hide the signal coming from the quantum dot.[12] We optically characterized the sample in top-view geometry via μ-PhotoLuminescence (PL) measurements at low temperature (~10K), under non-resonant conditions. A continuous power excitation laser emitting at 658nm was used for power dependence and exchange splitting measurements. A pulsed laser, with central wavelength at 656nm, a repetition rate of 40MHz and a pulse duration of 400ps, was employed for time-resolved investigations.

A preliminary analysis over a large number of single pyramidal QDs (~60) showed an average emission wavelength of the neutral exciton equal to ~853nm, with a standard deviation of 1.57nm. As already anticipated, this represents only a small redshift if compared to previously studied, nitrogen free, 0.5 nm thick $In_{0.25}Ga_{0.75}As$ /GaAs QDs.[9] More importantly power dependence measurements revealed an unexpected anti-binding bi-exciton energy. The assignment of the peaks was conducted on the base of combined power dependence, lifetime measurements and FSS characterization. In Fig.2 the dynamics exciton/bi-exciton is illustrated for five different excitation power levels. At weak excitation levels the pure exciton peak at lower energy is dominant; the bi-exciton peak at higher energy becomes visible and more intense as the excitation power increases. From this power dependence evolution a clear anti-binding behaviour can be observed. This tendency was systematically found on all the dots analyzed, and the *average* value of the anti-binding energy was found to be ~3.3meV.

Time-resolved measurements confirmed the nature of the bi-exciton and neutral exciton assignment: they systematically exhibited lifetimes which were respectively half the other, within a small variance, as summarized in Fig.3 (a). Typical excitons decayed with a time constant of 1-4ns, but a non negligible (<7%) amount of neutral excitonic states, although emitting in the very same wavelength range, presents surprisingly longer lifetimes, up to 7ns. In Fig.3 (b)-(c) representative emission decays of the pair exciton/bi-exciton emitted from two different pyramidal QDs are plotted in logarithmic scale. Correspondent lifetimes were extrapolated by fitting the data and they are indicated in the same figure with the exponential fit. Further analysis and more systematic characterizations are still ongoing to identify the cause of the lifetime values disparity.

Polarization dependent measurements[13] revealed extremely low values of FSS overall the sample, being the highest measured splitting 3.4μeV. Fig.4 shows an exemplifying (Lorentzian fit) spectrum of linearly polarized (vertical, trace in blue, triangles and horizontal, red curve with dots) neutral excitons and bi-excitons emitted from a single $In_{0.25}Ga_{0.75}As_{1-\varepsilon}N_{\varepsilon}$/GaAs QD. The differently polarized emission spectra are not distinguishable, suggesting that the FSS value is limited by the combined system and fitting procedure resolution (<3 μeV). Compared to our Nitrogen free QDs, the introduction of Nitrogen in the structure seems to suggest a remarkable improvement in terms of asymmetries of the crystallographic structure. Although the substrate orientation (111) and the patterning process assure a threefold $C_{3v}$ symmetry, ideally sufficient to guarantee a nearly zero excitonic anisotropic exchange splitting,[14] we always observe in our counterpart Nitrogen free samples a non negligible FSS (> 10μeV).[15]

In conclusion we have encountered a number of unforeseen optical features in $In_{0.25}Ga_{0.75}As_{1-\varepsilon}N_{\varepsilon}$/GaAs pyramidal site-controlled QD samples. While further investigations are needed to determine the exact incorporation and localization of Nitrogen atoms in the QD system (e.g. nonlinear incorporation of Nitrogen due to In segregation,[16] distribution of the atoms and their possible diffusion into the confining barriers),[17] the exposure to U-DMHydrazine during growth resulted in a significant change of the QD properties.

An anti-binding bi-exciton was observed: this leads us to consider an hypothetical engineering mechanism of the bi-exciton binding energy (from negative to positive), based on a modulation of the U-DMHydrazine flux during growth. Time-resolved measurements emphasized the presence of a significant percentage of dots with long lifetimes. Exchange splitting analysis revealed an extremely symmetrical nature of our QDs, being the FSS measured on different dots nearly zero. Causes of in-plane asymmetry reduction and/or strain redistribution need to be analyzed in detail. It has been proposed by pseudopotential calculations and confirmed by structural microscopy that the fine structure in MBE grown self-assembled QDs is mainly originated from a non uniform incorporation of Indium in the islands.[18] This suggests that even in the case of MOVPE growth the interplay between U-DMHy and other precursors might lead to a different incorporation dynamics of In atoms, changing the segregation effects in the dot layer. Nevertheless a fine control of the exchange splitting via tuning of the Nitrogen incorporation can be conceived and interesting future opportunities are open for the generation of polarized entangled photons,

This research was enabled by the Irish Higher Education Authority Program for Research in Third Level Institutions (2007-2011) via the INSPIRE programme, and by Science Foundation Ireland under grants 05/IN.1/I25 and 08/RFP/MTR1659. We are grateful to K. Thomas for his support with the MOVPE system.

**Figure Captions**

Fig. 1: (colour online) Atomic Force Microscopy image of a typical pyramidal QD scanned in cross-section. The sequence of the layers is numbered: 1. GaAs buffer; 2. $Al_{0.8}Ga_{0.2}As$ etch stop cladding; 3. $Al_{0.55}Ga_{0.45}As$ barriers; 4. GaAs barriers. The 1.5nm thick $In_{0.35}Ga_{0.65}As$ dot is also indicated.

Fig. 2: (color online ) Low temperature excitation power dependence evolution of excitons and bi-excitons emitted from a pyramidal $In_{0.25}Ga_{0.75}As_{1-\varepsilon}N_{\varepsilon}/GaAs$ QD. Exciton (X) and bi-exciton (2X) peaks are labelled and their dynamics show a clear anti-binding behaviour,.

Fig. 3: (color online) a) Exciton (X, red trace with triangles) and bi-exciton (2X, blue trace with dots) lifetimes measured for some representative InGaAsN/GaAs QDs at low temperature. For each dot excitons present a lifetime twice the biexciton one. (b)-(c) Exciton (red) and bi-exciton (blue) emission decays, with relative exponential fit curves (smooth continuous black lines), in logarithmic scale, as function of time measured from two different InGaAsN/GaAs QDs. Their lifetimes are indicated next to the correspondent curve.

Fig. 4: (color online) Lorentzian fit of normalized vertically (blue trace with triangular symbols) and horizontally (red trace, dot symbols) polarized emission spectra measured from a representative InGaAsN/GaAs QD. The two orthogonal

linearly polarized emissions overlap, indicating a nearly zero value of the FSS, limited by the set-up resolution (~2-3μeV).

**Figure 1**

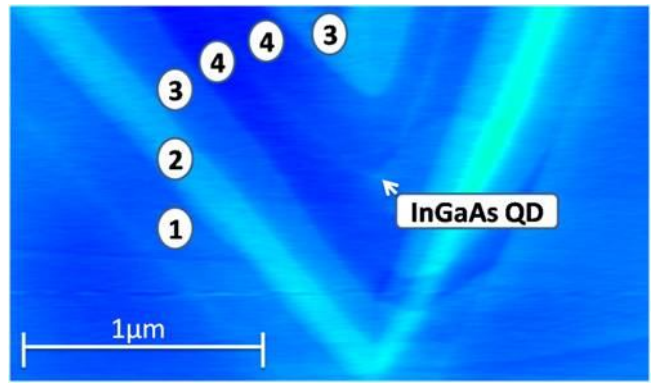

**Figure 2**

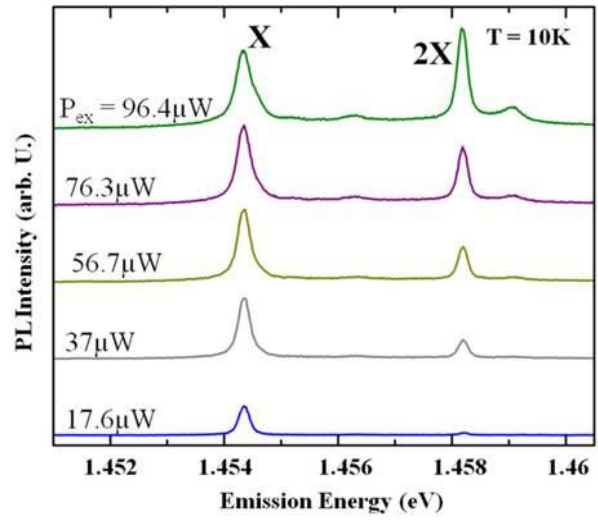

# Figure 3

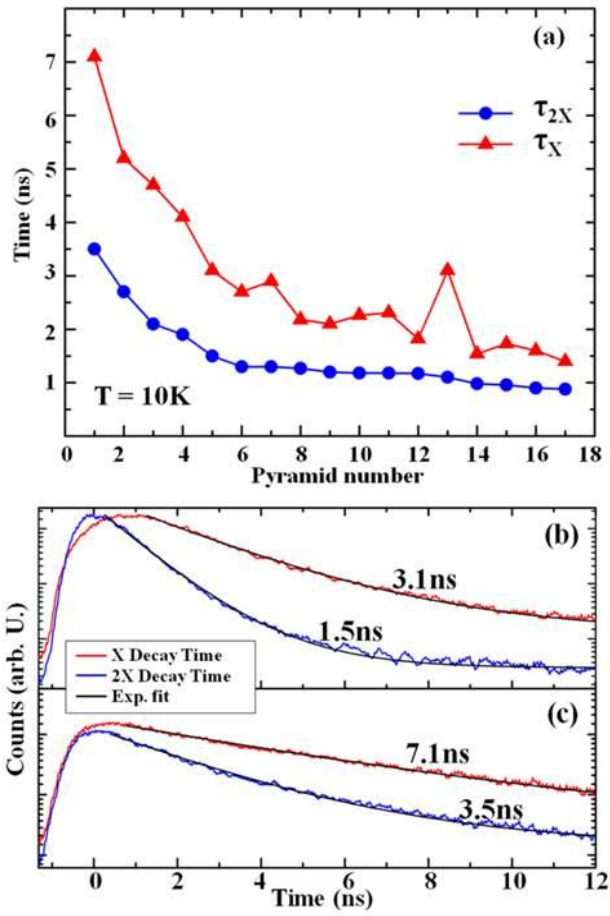

**Figure 4**

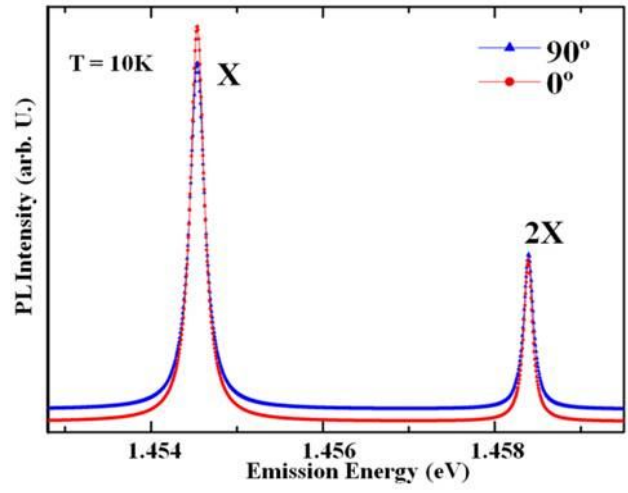